\documentclass[10pt, a4paper]{article}
\usepackage{float}
\usepackage{fullpage}
\usepackage{amsmath}
\usepackage{amsfonts}
\usepackage[ruled,vlined]{algorithm2e}
\usepackage{graphicx}
\usepackage{xcolor}
\usepackage{booktabs}
\usepackage{tikz}
\usetikzlibrary{shapes.geometric, positioning, arrows.meta, calc}

\usepackage[font=small,labelfont=bf]{caption}

\usepackage[
backend=biber,
style=alphabetic,
sorting=ynt
]{biblatex}
\usepackage[colorlinks=true, citecolor=blue]{hyperref}
\usepackage{url}

\SetAlgoNlRelativeSize{-1}

\tikzstyle{annotationText} = [font=\scriptsize\sffamily, text=black!70]
\tikzstyle{structureText} = [font=\large\sffamily, text=black!70]

\tikzstyle{generalLine} = [->, thick, black]
\tikzstyle{annotationLine} = [->, dashed, black]
\tikzstyle{structureLine} = [<->, thick, black!70]

\tikzstyle{highlightedNode} = [draw=blue!60!black, fill=blue!5, line width=1pt]
\tikzstyle{generalNode} = [draw, line width=1pt]
\tikzstyle{annotationNode} = [draw, dashed, text=black!70, shape=ellipse]
\tikzstyle{emptyNode} = [draw, dashed, line width=1pt, fill=gray!30]

\title{\textbf{Cartesian Merkle Tree}}

\author{Artem Chystiakov, Oleh Komendant, Kyrylo Riabov}

\date{April 2025}

\begin{document}
\maketitle

\begin{abstract}

This paper introduces the Cartesian Merkle Tree, a deterministic data structure that combines the properties of a Binary Search Tree, a Heap, and a Merkle tree. The Cartesian Merkle Tree supports insertions, updates, and removals of elements in $O(\log n)$ time, requires $n$ space, and enables membership and non-membership proofs via Merkle-based authentication paths. This structure is particularly suitable for zero-knowledge applications, blockchain systems, and other protocols that require efficient and verifiable data structures.

\end{abstract}

\section{Introduction}

Cartesian Merkle Tree (CMT) research is motivated by the increasing demand for efficient and secure data structures in cryptographic, blockchain, and zero-knowledge (ZK) systems. Traditional Merkle trees, such as Sparse Merkle Trees (SMT)\cite{sparse_merkle_tree}, are widely used but have limitations in terms of balance and efficiency.

CMTs integrate binary search tree properties for key-based organization and heap properties for priority balancing, data storage optimization, elements retrieval, and proof generation. To ensure determinism, the priority value for an element is derived from its key using a predefined algorithm, such as a hash function.

CMTs offer an efficient alternative to SMTs, reducing memory usage while preserving the key properties of the latter. One of the key features of CMT is that it stores useful data in every node, unlike SMTs which only store it in the leaves. The time complexity of the operations is still $O(\log n)$, with the only trade-off being a Merkle proof size at worst two times larger than SMTs.

\section{Background}

Merkle trees are fundamental to many cryptographic protocols, enabling efficient and secure proofs of data inclusion. They are widely used across various systems, ranging from maintaining validator sets in Ethereum to storing the state of zero-knowledge layer-2 rollups. The primary strength of Merkle trees lies in their ability to prove the inclusion of data in a highly compact and efficient manner.

Although the standard Merkle tree structure is binary, there are variations, such as the Merkle-Patricia tree, which utilizes sixteen child nodes per branch, optimizing performance in specific contexts. Despite these advances, Merkle trees still face challenges related to storage and operational complexity. Typically, they require $O(\log n)$ operations and $2n$ storage for binary trees.

While vanilla Merkle trees are often impractical for on-chain usage, they are still frequently utilized in off-chain whitelists and similar applications. However, two significant Merkle tree modifications have emerged that unlock the full potential of the data structure: Incremental Merkle Trees (IMT) and Sparse Merkle Trees (SMT).

\subsection{Incremental Merkle Tree}

IMT is a push-only data structure that can be used on-chain to build the tree, but requires an off-chain service to generate inclusion proofs. IMT is currently used by TornadoCash for anonymizing depositors, by Semaphore to store group membership commitments, and by the BeaconChain deposit smart contract to manage the list of Ethereum validators.

Unlike standard Merkle trees, IMTs do not store individual elements. Instead, they are merely used to ``build" the root. IMTs require $\log n$ storage, with inclusion proofs also having $O(\log n)$ complexity. Importantly, IMTs cannot be used independently without an off-chain service, as the entire tree must be reconstructed to generate an inclusion proof.

\subsection{Sparse Merkle Tree}

Until recently, SMTs were regarded as one of the most efficient data structures, particularly in the context of blockchain technologies and ZK systems. For instance, SMT is used by Scroll to maintain its ZK rollup state, by Rarimo to store ZK-provable national passport-based identity data, and by iden3 to manage custom-issued on-chain identities.

SMT is a particularly fascinating data structure. Unlike IMT, SMT does not require any off-chain services. It is deterministic, meaning the structure of the tree remains identical for a given set of elements, regardless of their insertion order. The position of an element in the tree is determined by its bitwise prefix: if a $0$ is encountered, the left child is selected; if a $1$ is encountered, the right child is chosen. This results in the tree size being constrained by the bitwise length of this prefix. However, reaching depths of $97$ or more is currently infeasible due to limitations within the EVM stack. In practice, the size of the tree \textit{without collisions} cannot exceed $2^{50}$, or approximately $1.12 \times 10^{15}$ (one quadrillion) elements, according to the Birthday Problem\cite{wikipedia-birthday}.

\section{Data Structure Description}

A Cartesian Merkle Tree can be seen as a standard Cartesian tree or Treap with the additional data Merkleization property. Each element in CMT corresponds to a point on a two-dimensional plane, with the key \texttt{k} representing the X-coordinate and the priority \texttt{p} representing the Y-coordinate. In traditional Cartesian trees \cite{wiki:treap}, the value of \texttt{p} is typically chosen at random, which contributes to a more balanced tree structure. However, if \texttt{p} is deterministically derived from \texttt{k}, then the same key will always produce the same point on the plane. As a result, the structure of the CMT becomes deterministic.

Let $\texttt{e} = \texttt{k}$ be a new entry in the tree \texttt{T}, where \texttt{k} is the key of the entry. The entry \texttt{e} may also optionally have a field \texttt{v}, the value. The node where this data element \texttt{e} is stored is determined based on the information contained in \texttt{e}. Let \texttt{H} be a cryptographically secure hash function that returns the hash result for an arbitrary number of values. Let $\texttt{p} = P_H({\texttt{e}})$ be the priority of the element \texttt{e}, where $P_H$ can be any deterministic algorithm that transforms the values of \texttt{e} into a number.

Let $\texttt{node} = (\texttt{k}, \texttt{p}, \texttt{mh})$ be any node in the tree \texttt{T}, where \texttt{mh} is the Merkle Hash value of the node, calculated as follows:
\[
mh = H(\texttt{entry} \parallel \texttt{leftChildMH} \parallel \texttt{rightChildMH})
\]
Here, $\texttt{leftChildMH}$ and $\texttt{rightChildMH}$ are the Merkle hash values of the node’s children, sorted in ascending order, and $\texttt{entry}$ is the node's useful payload. If a child is absent, its hash value is considered to be $0$.

The node where \texttt{e} should be stored is determined by the following rules:
\[
\texttt{leftChild.k} \leq \texttt{e.k} \leq \texttt{rightChild.k}
\]
\[
\texttt{leftChild.p} \leq \texttt{e.p} \quad \text{and} \quad \texttt{rightChild.p} \leq \texttt{e.p}
\]

The first rule ensures the binary search tree property, while the second rule maintains the min-heap property.

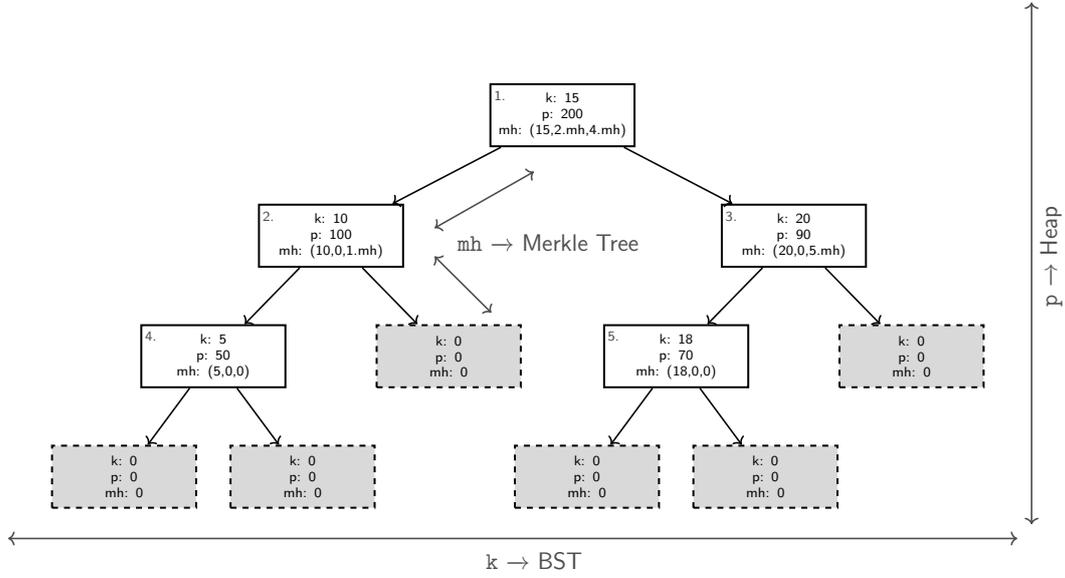
\begin{figure}[H]
    \centering
    \resizebox{0.9\textwidth}{!}{
        \begin{tikzpicture}[
            every node/.style={rectangle, text width=2.3cm, minimum height=1.1cm, align = center, font=\scriptsize\sffamily}
        ]
            \node (root) [generalNode] {k: 15 \\ p: 200 \\ mh: (15,2.mh,4.mh)};
            \node (left1) [generalNode, below left=of root, xshift=-0.5cm, yshift=0cm] {k: 10 \\ p: 100 \\ mh: (10,0,1.mh)};
            \node (right1) [generalNode, below right=of root, xshift=0.5cm, yshift=0cm] {k: 20 \\ p: 90 \\ mh: (20,0,5.mh)};

            \node (left1-1) [generalNode, below left=of left1, xshift=1.5cm, yshift=0cm] {k: 5 \\ p: 50 \\ mh: (5,0,0)};
            \node (left1-2) [emptyNode, below right=of left1, xshift=-1.5cm, yshift=0cm] {k: 0 \\ p: 0 \\ mh: 0};

            \node (left2-1) [emptyNode, below left=of left1-1, xshift=2cm, yshift=0cm] {k: 0 \\ p: 0 \\ mh: 0};
            \node (left2-2) [emptyNode, below right=of left1-1, xshift=-2cm, yshift=0cm] {k: 0 \\ p: 0 \\ mh: 0};

            \node (right1-1) [generalNode, below left=of right1, xshift=1.5cm, yshift=0cm] {k: 18 \\ p: 70 \\ mh: (18,0,0)};
            \node (right1-2) [emptyNode, below right=of right1, xshift=-1.5cm, yshift=0cm] {k: 0 \\ p: 0 \\ mh: 0};

            \node (right2-1) [emptyNode, below left=of right1-1, xshift=2cm, yshift=0cm] {k: 0 \\ p: 0 \\ mh: 0};
            \node (right2-2) [emptyNode, below right=of right1-1, xshift=-2cm, yshift=0cm] {k: 0 \\ p: 0 \\ mh: 0};

            \node[annotationText, xshift=-1.1cm, yshift=0.35cm] at (root) {1.};
            \node[annotationText, xshift=-1.1cm, yshift=0.35cm] at (right1) {3.};
            \node[annotationText, xshift=-1.1cm, yshift=0.35cm] at (left1) {2.};
            \node[annotationText, xshift=-1.1cm, yshift=0.35cm] at (left1-1) {4.};
            \node[annotationText, xshift=-1.1cm, yshift=0.35cm] at (right1-1) {5.};

            \draw [structureLine] (-9.75cm, -7.5cm) -- (8cm, -7.5cm);
            \draw [structureLine] (8.25cm, 2cm) -- (8.25cm, -7.25cm);
            \draw [structureLine] (-0.5cm, -1cm) -- (-2.25cm, -2cm);
            \draw [structureLine] (-2.25cm, -2.5cm) -- (-1.25cm, -3.5cm);

            \node (bst-annotation) [structureText, xshift=0cm, yshift=-0.35cm] at (-0.5cm, -7.55cm) {\texttt{k} $\xrightarrow{}$ BST};
            \node (heap-annotation) [structureText, xshift=0.3cm, yshift=-0cm, rotate=90] at (8.3cm, -2.5cm) {\texttt{p} $\xrightarrow{}$ Heap};
            \node (merkle-tree-annotation) [structureText, text width=4cm, xshift=1cm, yshift=-0cm] at (-1.25cm, -2.25cm) {\texttt{mh} $\xrightarrow{}$ Merkle Tree};

            \draw[generalLine] (root) -- (left1);
            \draw[generalLine] (root) -- (right1);
            \draw[generalLine] (left1) -- (left1-1);
            \draw[generalLine] (left1) -- (left1-2);
            \draw[generalLine] (left1-1) -- (left2-1);
            \draw[generalLine] (left1-1) -- (left2-2);
            \draw[generalLine] (right1) -- (right1-1);
            \draw[generalLine] (right1) -- (right1-2);
            \draw[generalLine] (right1-1) -- (right2-1);
            \draw[generalLine] (right1-1) -- (right2-2);
        \end{tikzpicture}
    }
    \caption{Example of CMT}
    \label{fig:example-cmt}
\end{figure}

\begin{algorithm}[H]
\caption{CMT Utils Functions}
\label{alg:utils}

\SetKwFunction{FnMH}{CalculateMH}
\SetKwFunction{FnRR}{RightRotate}
\SetKwFunction{FnLR}{LeftRotate}
\SetKwProg{Fn}{Function}{:}{}

\Fn{\FnMH{\texttt{key}, \texttt{leftChildMH}, \texttt{rightChildMH}}}{
    \textcolor{green!40!black!70}{\tcc{Sort leftChildMH, rightChildMH in ascending order}}
    \If{$\texttt{leftChildMH} < \texttt{rightChildMH}$}{
        \Return{H($\texttt{key} || \texttt{leftChildMH} || \texttt{rightChildMH}$)}\;
    }
    \Else{
        \Return{H($\texttt{key} || \texttt{rightChildMH} || \texttt{leftChildMH}$)}\;
    }
}\

\Fn{\FnRR{\texttt{node}}}{
    \textcolor{green!40!black!70}{\tcc{Save pointers to the left child and its right child}}
    \texttt{currentLeftChild} $\gets$ \texttt{node.leftChild}\;
    \texttt{newLeftChild} $\gets$ \texttt{currentLeftChild.rightChild}\;

    \textcolor{green!40!black!70}{\tcc{Perform the right rotation by updating the pointers}}
    \texttt{currentLeftChild.rightChild} $\gets$ \texttt{node}\;
    \texttt{node.leftChild} $\gets$ \texttt{newLeftChild}\;

    \textcolor{green!40!black!70}{\tcc{Update the mh value of the node after rotation}}
    \texttt{node.mh} $\gets$ \FnMH{\texttt{node.e.k}, \texttt{leftChildMH}, \texttt{rightChildMH}}\;
}\

\Fn{\FnLR{\texttt{node}}}{
    \textcolor{green!40!black!70}{\tcc{Save pointers to the right child and its left child}}
    \texttt{currentRightChild} $\gets$ \texttt{node.rightChild}\;
    \texttt{newRightChild} $\gets$ \texttt{currentRightChild.leftChild}\;

    \textcolor{green!40!black!70}{\tcc{Perform the left rotation by updating the pointers}}
    \texttt{currentRightChild.leftChild} $\gets$ \texttt{node}\;
    \texttt{node.rightChild} $\gets$ \texttt{newRightChild}\;

    \textcolor{green!40!black!70}{\tcc{Update the mh value of the node after rotation}}
    \texttt{node.mh} $\gets$ \FnMH{\texttt{node.e.k}, \texttt{leftChildMH}, \texttt{rightChildMH}}\;
}

\end{algorithm}

\subsection{Insertion}

When inserting an entry \texttt{e}, the corresponding node \texttt{n} in the tree \texttt{T} is determined by traversing downward from the root while maintaining the BST property. Once inserted, the structure is recursively adjusted upward to the root to restore the min-heap property using left or right rotations. At each modification of any node, the \texttt{mh} value is also recomputed.

\vspace{0.5cm}

\begin{algorithm}[H]
\caption{Insertion of an Element into the CMT}
\label{alg:insert}
\KwIn{Element $e = k$ to be inserted}
\KwOut{Updated tree \texttt{T}}

\SetKwFunction{Insert}{Insert}
\SetKwProg{Fn}{Function}{:}{}

\Fn{\Insert{$T, e$}} {
    Find the appropriate position for \texttt{e} in \texttt{T} based on the BST property\;
    Create a new node \texttt{n} to insert \texttt{e}, and set:\\
    \Begin{
        Set $\texttt{n.e} \gets \texttt{e}$\;
        Set $\texttt{n.p} \gets P_H(\texttt{e})$\;
        Set $\texttt{n.mh} \gets \FnMH{\texttt{e.k}, \texttt{leftChildMH}, \texttt{rightChildMH}}$\;
    }
    \textcolor{green!40!black!70}{\tcc{Restore min-heap property by rotating upwards}}
    \texttt{Node} $\gets$ \texttt{n}\;
    \While{\texttt{Node} is not root and $\texttt{Parent.Priority} < \texttt{Node.Priority}$}{
        \textcolor{green!40!black!70}{\tcc{Determine which child the node is to its parent and perform needed rotation}}
        \If{$\texttt{Parent.leftChild} = \texttt{Node}$}{
            \FnRR{\texttt{Parent}}\;
        }
        \ElseIf{$\texttt{Parent.rightChild} = \texttt{Node}$}{
            \FnLR{\texttt{Parent}}\;
        }
        \texttt{Node} $\gets$ \texttt{Parent(Node)}\;
        \textcolor{green!40!black!70}{\tcc{Update mh value of the node after rotation}}
        $\texttt{Node.mh} \gets \FnMH{\texttt{Node.e.k}, \texttt{leftChildMH}, \texttt{rightChildMH}}$\;
    }
}

\end{algorithm}

\vspace{0.5cm}
\noindent
\underline{\large Example}

\vspace{0.5cm}
\noindent

In this example, consider the insertion of the element \texttt{e} where $\texttt{e.k} = 13$ and $P_H(\texttt{e}) = 250$ into the tree shown in {Figure~\ref{fig:example-cmt}}. After the element insertion, the tree will have the following structure:

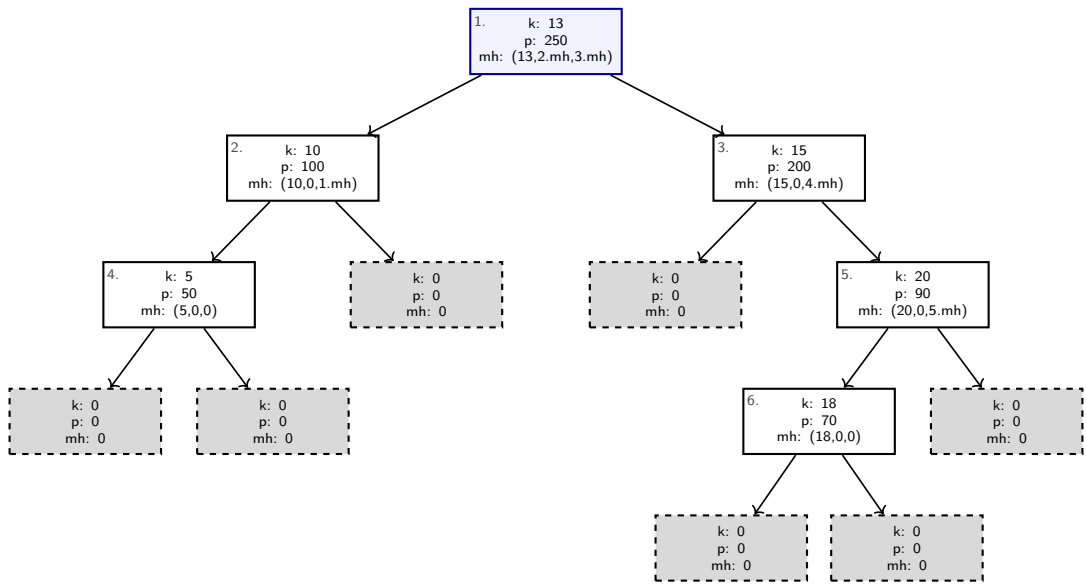
\begin{figure}[H]
    \centering
    \resizebox{0.9\textwidth}{!}{
        \begin{tikzpicture}[
            every node/.style={rectangle, text width=2.3cm, minimum height=1.1cm, align = center, font=\scriptsize\sffamily}
        ]
            \node (root) [highlightedNode] {k: 13 \\ p: 250 \\ mh: (13,2.mh,3.mh)};

            \node (left1) [generalNode, below left=of root, xshift=-0.5cm, yshift=0cm] {k: 10 \\ p: 100 \\ mh: (10,0,1.mh)};
            \node (right1) [generalNode, below right=of root, xshift=0.5cm, yshift=0cm] {k: 15 \\ p: 200 \\ mh: (15,0,4.mh)};

            \node (left1-1) [generalNode, below left=of left1, xshift=1.5cm, yshift=0cm] {k: 5 \\ p: 50 \\ mh: (5,0,0)};
            \node (left1-2) [emptyNode, below right=of left1, xshift=-1.5cm, yshift=0cm] {k: 0 \\ p: 0 \\ mh: 0};

            \node (right1-1) [emptyNode, below left=of right1, xshift=1.5cm, yshift=0cm] {k: 0 \\ p: 0 \\ mh: 0};
            \node (right1-2) [generalNode, below right=of right1, xshift=-1.5cm, yshift=0cm] {k: 20 \\ p: 90 \\ mh: (20,0,5.mh)};

            \node (left2-1) [emptyNode, below left=of left1-1, xshift=2cm, yshift=0cm] {k: 0 \\ p: 0 \\ mh: 0};
            \node (left2-2) [emptyNode, below right=of left1-1, xshift=-2cm, yshift=0cm] {k: 0 \\ p: 0 \\ mh: 0};

            \node (right2-1) [generalNode, below left=of right1-2, xshift=2cm, yshift=0cm] {k: 18 \\ p: 70 \\ mh: (18,0,0)};
            \node (right2-2) [emptyNode, below right=of right1-2, xshift=-2cm, yshift=0cm] {k: 0 \\ p: 0 \\ mh: 0};

            \node (right3-1) [emptyNode, below left=of right2-1, xshift=2.1cm, yshift=0cm] {k: 0 \\ p: 0 \\ mh: 0};
            \node (right3-2) [emptyNode, below right=of right2-1, xshift=-2.1cm, yshift=0cm] {k: 0 \\ p: 0 \\ mh: 0};

            \node[annotationText, xshift=-1.1cm, yshift=0.35cm] at (root) {1.};
            \node[annotationText, xshift=-1.1cm, yshift=0.35cm] at (right1) {3.};
            \node[annotationText, xshift=-1.1cm, yshift=0.35cm] at (left1) {2.};
            \node[annotationText, xshift=-1.1cm, yshift=0.35cm] at (left1-1) {4.};
            \node[annotationText, xshift=-1.1cm, yshift=0.35cm] at (right1-2) {5.};
            \node[annotationText, xshift=-1.1cm, yshift=0.35cm] at (right2-1) {6.};

            \draw[generalLine] (root) -- (left1);
            \draw[generalLine] (root) -- (right1);
            \draw[generalLine] (left1) -- (left1-1);
            \draw[generalLine] (left1) -- (left1-2);
            \draw[generalLine] (left1-1) -- (left2-1);
            \draw[generalLine] (left1-1) -- (left2-2);
            \draw[generalLine] (right1) -- (right1-1);
            \draw[generalLine] (right1) -- (right1-2);
            \draw[generalLine] (right1-2) -- (right2-1);
            \draw[generalLine] (right1-2) -- (right2-2);
            \draw[generalLine] (right2-1) -- (right3-1);
            \draw[generalLine] (right2-1) -- (right3-2);
        \end{tikzpicture}
    }
    \caption{CMT after insertion}
    \label{fig:insert-cmt}
\end{figure}

\subsection{Removal}

To remove an entry \texttt{e}, first determine whether there exists a node \texttt{n} in the tree \texttt{T} that corresponds to \texttt{e}. If such a node is found, assign $\texttt{n.p} = -\infty$, ensuring that during the heap property restoration via rotations, \texttt{n} will be moved to a leaf position. Once the node becomes a leaf, it can be easily removed. After removal, recursively traverse upward and update the \texttt{mh} values.

\vspace{0.5cm}

\begin{algorithm}[H]
\caption{Removal of an Element from the CMT}
\label{alg:remove}
\KwIn{Element $e = k$ to be removed}
\KwOut{Updated tree \texttt{T}}

\SetKwFunction{Remove}{Remove}
\SetKwFunction{FnMH}{CalculateMH}
\SetKwProg{Fn}{Function}{:}{}

\Fn{\Remove{$T, e$}} {
    Find the node \texttt{n} corresponding to \texttt{e} in \texttt{T}\;
    \If{\texttt{n} does not exist}{
        \textbf{Throw error:} ``Element not found"\;
    }

    \textcolor{green!40!black!70}{\tcc{Mark node for removal by setting its priority to $-\infty$}}
    \texttt{n.p} $\gets -\infty$\;

    \textcolor{green!40!black!70}{\tcc{Restore min-heap property via rotations}}
    \texttt{Node} $\gets$ \texttt{n}\;
    \While{\texttt{Node} is not a leaf}{
        \textcolor{green!40!black!70}{\tcc{Determine which child has higher priority and perform rotation}}
        \If{$\texttt{Node.leftChild.p} > \texttt{Node.rightChild.p}$}{
            \FnRR{\texttt{Node}}\;
        }
        \Else{
            \FnLR{\texttt{Node}}\;
        }
        \texttt{Node} $\gets$ new position after rotation\;
    }

    \textcolor{green!40!black!70}{\tcc{Remove the node}}
    Remove \texttt{Node} from \texttt{T}\;

    \textcolor{green!40!black!70}{\tcc{Update \texttt{mh} values while traversing upwards}}
    \texttt{Parent} $\gets$ \texttt{Parent(Node)}\;
    \While{\texttt{Parent} is not null}{
        $\texttt{Parent.mh} \gets \FnMH{\texttt{Parent.e.k}, \texttt{Parent.leftChildMH}, \texttt{Parent.rightChildMH}}$\;
        \texttt{Parent} $\gets$ \texttt{Parent(Parent)}\;
    }
}

\end{algorithm}

\vspace{0.5cm}
\noindent
\underline{\large Example}

\vspace{0.5cm}
\noindent

Consider removing of an element \texttt{e} with a key $\texttt{e.k} = 15$ from the tree shown in Figure~\ref{fig:insert-cmt}. After removing, the tree structure will be modified as follows:

\begin{figure}[H]
    \centering
    \resizebox{0.9\textwidth}{!}{
        \begin{tikzpicture}[
            every node/.style={rectangle, text width=2.3cm, minimum height=1.1cm, align = center, font=\scriptsize\sffamily}
        ]
            \node (root) [generalNode] {k: 13 \\ p: 250 \\ mh: (13,2.mh,3.mh)};

            \node (left1) [generalNode, below left=of root, xshift=-0.5cm, yshift=0cm] {k: 10 \\ p: 100 \\ mh: (10,0,1.mh)};
            \node (right1) [generalNode, below right=of root, xshift=0.5cm, yshift=0cm] {k: 20 \\ p: 90 \\ mh: (20,0,5.mh)};

            \node (left1-1) [generalNode, below left=of left1, xshift=1.5cm, yshift=0cm] {k: 5 \\ p: 50 \\ mh: (5,0,0)};
            \node (left1-2) [emptyNode, below right=of left1, xshift=-1.5cm, yshift=0cm] {k: 0 \\ p: 0 \\ mh: 0};

            \node (right1-1) [generalNode, below left=of right1, xshift=1.5cm, yshift=0cm] {k: 18 \\ p: 70 \\ mh: (18,0,0)};
            \node (right1-2) [emptyNode, below right=of right1, xshift=-1.5cm, yshift=0cm] {k: 0 \\ p: 0 \\ mh: 0};

            \node (left2-1) [emptyNode, below left=of left1-1, xshift=2cm, yshift=0cm] {k: 0 \\ p: 0 \\ mh: 0};
            \node (left2-2) [emptyNode, below right=of left1-1, xshift=-2cm, yshift=0cm] {k: 0 \\ p: 0 \\ mh: 0};

            \node (right2-1) [emptyNode, below left=of right1-1, xshift=2cm, yshift=0cm] {k: 0 \\ p: 0 \\ mh: 0};
            \node (right2-2) [emptyNode, below right=of right1-1, xshift=-2cm, yshift=0cm] {k: 0 \\ p: 0 \\ mh: 0};

            \node[annotationText, xshift=-1.1cm, yshift=0.35cm] at (root) {1.};
            \node[annotationText, xshift=-1.1cm, yshift=0.35cm] at (right1) {3.};
            \node[annotationText, xshift=-1.1cm, yshift=0.35cm] at (left1) {2.};
            \node[annotationText, xshift=-1.1cm, yshift=0.35cm] at (left1-1) {4.};
            \node[annotationText, xshift=-1.1cm, yshift=0.35cm] at (right1-1) {5.};

            \draw[generalLine] (root) -- (left1);
            \draw[generalLine] (root) -- (right1);
            \draw[generalLine] (left1) -- (left1-1);
            \draw[generalLine] (left1) -- (left1-2);
            \draw[generalLine] (left1-1) -- (left2-1);
            \draw[generalLine] (left1-1) -- (left2-2);
            \draw[generalLine] (right1) -- (right1-1);
            \draw[generalLine] (right1) -- (right1-2);
            \draw[generalLine] (right1-1) -- (right2-1);
            \draw[generalLine] (right1-1) -- (right2-2);
        \end{tikzpicture}
    }
    \caption{CMT after removal}
    \label{fig:remove-cmt}
\end{figure}
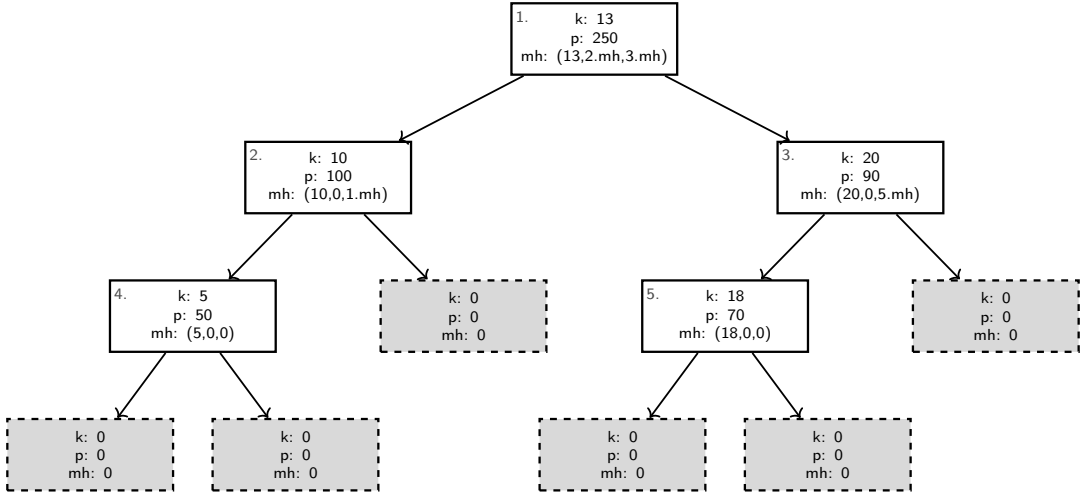

\subsection{CMT Proof}

Let \texttt{CMT proof} be a proof of membership of an element \texttt{e} in a tree \texttt{T}, represented as $\texttt{proof} = [\texttt{prefix}, \texttt{suffix}]$, where:
\begin{itemize}
    \item \texttt{prefix} is an ordered list of Merkle path nodes, each containing pairs of values $(\texttt{n.e.k}, \texttt{n.mh})$ for each node \texttt{n};
    \item \texttt{suffix} consists of \texttt{e.leftChildMH} and \texttt{e.rightChildMH}, representing the subtree structure of \texttt{e};
    \item \texttt{existence} is a boolean flag indicating whether \texttt{e} exists in the tree;
    \item \texttt{nonExistenceKey} is used when \texttt{e} does not exist in the tree, and helps verify that \texttt{e} is absent.
\end{itemize}

The initial value of \texttt{acc} is computed as:
\[
\texttt{acc} = H((\texttt{existence} ? \texttt{e.k} : \texttt{nonExistenceKey}) \parallel \texttt{proof.suffix[0]} \parallel \texttt{proof.suffix[1]})
\]
ensuring that $\texttt{proof.suffix[0]} < \texttt{proof.suffix[1]}$.

Then, \texttt{acc} is iteratively updated using values from \texttt{prefix}:
\[
\begin{cases}
\texttt{acc} = H(\texttt{n.e.k} \parallel \texttt{n.mh} \parallel \texttt{acc}), & \text{if } \texttt{n.mh} < \texttt{acc}, \\
\texttt{acc} = H(\texttt{n.e.k} \parallel \texttt{acc} \parallel \texttt{n.mh}), & \text{otherwise}.
\end{cases}
\]

The proof is considered valid if the final value of \texttt{acc} matches the root of \texttt{T}.

\begin{algorithm}[H]
\caption{CMT Proof Generation}
\label{alg:cmt_proof_generation}
\KwIn{Element $e = k$ to be proven in tree \texttt{T}}
\KwOut{Proof $\texttt{proof} = [\texttt{prefix}, \texttt{suffix}, \texttt{existence}, \texttt{nonExistenceKey}]$}

\SetKwFunction{GenerateProof}{GenerateProof}
\SetKwProg{Fn}{Function}{:}{}

\Fn{\GenerateProof{$T, e$}} {
    Initialize empty lists: $\texttt{prefix} \gets []$, $\texttt{suffix} \gets []$\;
    Initialize bool variable $\texttt{existence} \gets$ true\;
    Initialize variable $\texttt{currentNode} \gets$ null\;

    \textcolor{green!40!black!70}{\tcc{Get the appropriate node for the entry \texttt{e}}}
    \If{\texttt{e} does not exist in the tree \texttt{T}}{
        $\texttt{currentNode} \gets$ node with appropriate key for non-existence proof\;
        $\texttt{existence} \gets$ \texttt{false}\;
        $\texttt{nonExistenceKey} \gets \texttt{currentNode.e.k}$\;
    } \Else{
        $\texttt{currentNode} \gets$ node in \texttt{T} where $\texttt{n.e} = \texttt{e}$\;
    }

    \textcolor{green!40!black!70}{\tcc{Set \texttt{suffix} as the hash values of \texttt{currentNode}'s children}}
    $\texttt{suffix} \gets [\texttt{n.leftChildMH}, \texttt{n.rightChildMH}]$\;

    \textcolor{green!40!black!70}{\tcc{Construct \texttt{prefix} by traversing the path to the root}}
    \While{\texttt{currentNode} is not root}{
        $\texttt{parent} \gets \texttt{Parent(currentNode)}$\;
        Append $(\texttt{parent.e.k}, \texttt{parent.mh})$ to \texttt{prefix}\;
        $\texttt{currentNode} \gets \texttt{parent}$\;
    }

    \Return $[\texttt{prefix}, \texttt{suffix}, \texttt{existence}, \texttt{nonExistenceKey}]$\;
}
\end{algorithm}

\vspace{0.5cm}

\begin{algorithm}[H]
\caption{Verification of CMT Proof}
\label{alg:verify_cmt_proof}
\KwIn{Proof $\texttt{proof} = [\texttt{prefix}, \texttt{suffix}, \texttt{existence}, \texttt{nonExistenceKey}]$, Element \texttt{e}, Root hash \texttt{root}}
\KwOut{\textbf{true} if \texttt{e} is in the tree or \texttt{e} is not in the tree and \texttt{existence} is false, \textbf{false} otherwise}

\SetKwFunction{VerifyProof}{VerifyProof}
\SetKwProg{Fn}{Function}{:}{}

\Fn{\VerifyProof{$proof, e, root$}}{
    \textcolor{green!40!black!70}{\tcc{Initialize \texttt{acc} with the element's Merkle hash}}
    \If{\texttt{proof.existence}}{
        $\texttt{acc} \gets \FnMH(\texttt{e.k}, \texttt{proof.suffix[0]}, \texttt{proof.suffix[1]})$\;
    } \Else{
        $\texttt{acc} \gets \FnMH(\texttt{nonExistenceKey}, \texttt{proof.suffix[0]}, \texttt{proof.suffix[1]})$\;
    }

    \textcolor{green!40!black!70}{\tcc{Iteratively compute the hash up the Merkle path}}
    \ForEach{$(\texttt{k}, \texttt{mh})$ in \texttt{proof.prefix}}{
        $\texttt{acc} \gets \FnMH(\texttt{k}, \texttt{acc}, \texttt{mh})$\;
    }

    \textcolor{green!40!black!70}{\tcc{Check if computed hash matches the root}}
    \Return $\texttt{acc} = \texttt{root}$\;
}
\end{algorithm}

\vspace{0.5cm}
\noindent
\underline{\large Example}

\vspace{0.5cm}
\noindent

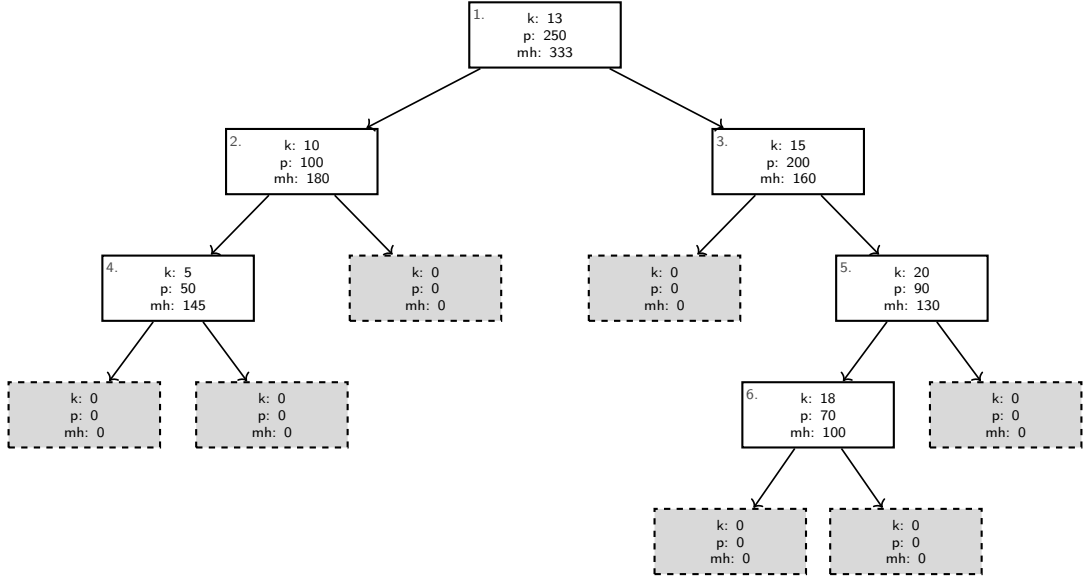
\begin{figure}[H]
    \centering
    \resizebox{0.9\textwidth}{!}{
        \begin{tikzpicture}[
            every node/.style={rectangle, text width=2.3cm, minimum height=1.1cm, align = center, font=\scriptsize\sffamily}
        ]
            \node (root) [generalNode] {k: 13 \\ p: 250 \\ mh: 333};

            \node (left1) [generalNode, below left=of root, xshift=-0.5cm, yshift=0cm] {k: 10 \\ p: 100 \\ mh: 180};
            \node (right1) [generalNode, below right=of root, xshift=0.5cm, yshift=0cm] {k: 15 \\ p: 200 \\ mh: 160};

            \node (left1-1) [generalNode, below left=of left1, xshift=1.5cm, yshift=0cm] {k: 5 \\ p: 50 \\ mh: 145};
            \node (left1-2) [emptyNode, below right=of left1, xshift=-1.5cm, yshift=0cm] {k: 0 \\ p: 0 \\ mh: 0};

            \node (right1-1) [emptyNode, below left=of right1, xshift=1.5cm, yshift=0cm] {k: 0 \\ p: 0 \\ mh: 0};
            \node (right1-2) [generalNode, below right=of right1, xshift=-1.5cm, yshift=0cm] {k: 20 \\ p: 90 \\ mh: 130};

            \node (left2-1) [emptyNode, below left=of left1-1, xshift=2cm, yshift=0cm] {k: 0 \\ p: 0 \\ mh: 0};
            \node (left2-2) [emptyNode, below right=of left1-1, xshift=-2cm, yshift=0cm] {k: 0 \\ p: 0 \\ mh: 0};

            \node (right2-1) [generalNode, below left=of right1-2, xshift=2cm, yshift=0cm] {k: 18 \\ p: 70 \\ mh: 100};
            \node (right2-2) [emptyNode, below right=of right1-2, xshift=-2cm, yshift=0cm] {k: 0 \\ p: 0 \\ mh: 0};

            \node (right3-1) [emptyNode, below left=of right2-1, xshift=2.1cm, yshift=0cm] {k: 0 \\ p: 0 \\ mh: 0};
            \node (right3-2) [emptyNode, below right=of right2-1, xshift=-2.1cm, yshift=0cm] {k: 0 \\ p: 0 \\ mh: 0};

            \node[annotationText, xshift=-1.1cm, yshift=0.35cm] at (root) {1.};
            \node[annotationText, xshift=-1.1cm, yshift=0.35cm] at (right1) {3.};
            \node[annotationText, xshift=-1.1cm, yshift=0.35cm] at (left1) {2.};
            \node[annotationText, xshift=-1.1cm, yshift=0.35cm] at (left1-1) {4.};
            \node[annotationText, xshift=-1.1cm, yshift=0.35cm] at (right1-2) {5.};
            \node[annotationText, xshift=-1.1cm, yshift=0.35cm] at (right2-1) {6.};

            \draw[generalLine] (root) -- (left1);
            \draw[generalLine] (root) -- (right1);
            \draw[generalLine] (left1) -- (left1-1);
            \draw[generalLine] (left1) -- (left1-2);
            \draw[generalLine] (left1-1) -- (left2-1);
            \draw[generalLine] (left1-1) -- (left2-2);
            \draw[generalLine] (right1) -- (right1-1);
            \draw[generalLine] (right1) -- (right1-2);
            \draw[generalLine] (right1-2) -- (right2-1);
            \draw[generalLine] (right1-2) -- (right2-2);
            \draw[generalLine] (right2-1) -- (right3-1);
            \draw[generalLine] (right2-1) -- (right3-2);
        \end{tikzpicture}
    }
    \caption{CMT proof example}
    \label{fig:proof-cmt}
\end{figure}

Consider the generation and verification of a CMT proof for an entry \texttt{e}, where $\texttt{e.k} = 18$, in a tree depicted in Figure~\ref{fig:proof-cmt}. To make the example clearer, we replace the \texttt{mh} in all nodes with particular numbers.

\begin{algorithm}[H]
\caption{CMT Inclusion Proof Example}
\label{alg:cmt_proof_example1}
\KwIn{
    \begin{enumerate}
        \item \texttt{proof}: \begin{itemize}
            \item \texttt{prefix}: [$13, 180, 15, 0, 20, 0$]
            \item \texttt{suffix}: [$0, 0$]
            \item \texttt{existence}: true
            \item \texttt{nonExistenceKey}: $0$
        \end{itemize}
        \item \texttt{rootNodeMH}: $333$
    \end{enumerate}
}

\textcolor{green!40!black!70}{\tcc{Initialize \texttt{acc} with the element's Merkle hash}}
$\texttt{acc} = H(\texttt{e.k} \parallel \texttt{proof.suffix[0]} \parallel \texttt{proof.suffix[1]})$; \textcolor{green!40!black!70}{\tcc{$H(18 \parallel 0 \parallel 0) = 100$}}\

\textcolor{green!40!black!70}{\tcc{Compute the final \texttt{acc} value using the entries from the prefix.}}
$\texttt{acc} = H(\texttt{proof.prefix[4]} \parallel \texttt{proof.prefix[5]} \parallel \texttt{acc})$; \textcolor{green!40!black!70}{\tcc{$H(20 \parallel 0 \parallel 100) = 130$}}

$\texttt{acc} = H(\texttt{proof.prefix[2]} \parallel \texttt{proof.prefix[3]} \parallel \texttt{acc})$; \textcolor{green!40!black!70}{\tcc{$H(15 \parallel 0 \parallel 130) = 160$}}

$\texttt{acc} = H(\texttt{proof.prefix[0]} \parallel \texttt{acc} \parallel \texttt{proof.prefix[1]})$; \textcolor{green!40!black!70}{\tcc{$H(13 \parallel 160 \parallel 180) = 333$}}\

\textcolor{green!40!black!70}{\tcc{Compare \texttt{acc} with \texttt{rootNodeMH}}}
$\texttt{acc} == \texttt{root.mh}$; \textcolor{green!40!black!70}{\tcc{Result is \texttt{true}}}

\end{algorithm}

\vspace{0.5cm}

Consider the case where an entry \texttt{e} with key $\texttt{e.k} = 25$ is not present in the tree depicted in Figure~\ref{fig:proof-cmt}. In this scenario, the proof will be structured as follows:

\begin{algorithm}[H]
\caption{CMT Exclusion Proof Example}
\label{alg:cmt_proof_example2}
\KwIn{
    \begin{enumerate}
        \item \texttt{proof}: \begin{itemize}
            \item \texttt{prefix}: [$13, 180, 15, 0$]
            \item \texttt{suffix}: [$100, 0$]
            \item \texttt{existence}: false
            \item \texttt{nonExistenceKey}: $20$
        \end{itemize}
        \item \texttt{rootNodeMH}: $333$
    \end{enumerate}
}

\textcolor{green!40!black!70}{\tcc{Initialize \texttt{acc} with the element's Merkle hash}}
$\texttt{acc} = H(\texttt{proof.nonExistenceKey} \parallel \texttt{proof.suffix[0]} \parallel \texttt{proof.suffix[1]})$; \textcolor{green!40!black!70}{\tcc{$H(20 \parallel 0 \parallel 100) = 130$}}\

\textcolor{green!40!black!70}{\tcc{Compute the final \texttt{acc} value using the entries from the \texttt{prefix}.}}
$\texttt{acc} = H(\texttt{proof.prefix[2]} \parallel \texttt{proof.prefix[3]} \parallel \texttt{acc})$; \textcolor{green!40!black!70}{\tcc{$H(20 \parallel 0 \parallel 130) = 160$}}
$\texttt{acc} = H(\texttt{proof.prefix[0]} \parallel \texttt{acc} \parallel \texttt{proof.prefix[1]})$; \textcolor{green!40!black!70}{\tcc{$H(13 \parallel 160 \parallel 180) = 333$}}\

\textcolor{green!40!black!70}{\tcc{Compare \texttt{acc} with \texttt{rootNodeMH}}}
$\texttt{acc} == \texttt{root.mh}$; \textcolor{green!40!black!70}{\tcc{Result is \texttt{true}}}

\end{algorithm}

\section{Reference Implementation}

To provide a practical implementation of the Cartesian Merkle Tree and its proof verification, we reference two existing implementations in Solidity and Circom:

\begin{itemize}
    \item The Solidity implementation, available in the \texttt{solidity-lib} repository\cite{solidity-lib-cmt}, provides smart contract functionalities for CMT construction and proof generation.
    \item The Circom implementation, available in the \texttt{circom-lib} repository\cite{circom-lib}, offers a zk-SNARK-friendly circuit for verifying CMT proofs within zero-knowledge proofs.
\end{itemize}

These implementations are fully compatible: proofs generated by the Solidity implementation can be used to generate and verify zero-knowledge proofs in the Circom implementation.

\section{Benchmarks}

This section presents the benchmarking results for the \texttt{Insert} and \texttt{Remove} functions in the Solidity CMT implementation \cite{solidity-lib-cmt} using the \texttt{Keccak256} and \texttt{Poseidon} hash functions. A comparative analysis of EVM gas costs was performed for each function using different datasets. The tests were conducted with 100, 1000, 5000, and 10000 elements to show how data size impacts performance. The EVM gas costs can be considered as normalized computation units, therefore, if the operation takes more gas, it is more computationally intensive.

In order to ensure a fair comparison of the CMT and SMT structures benchmarks, the Solidity version of the SMT\cite{solidity-lib-smt} was tested using the same methods as the CMT. To maintain consistency, the \texttt{value} field was removed from the SMT \texttt{Node} structure so that they occupy the same number of storage slots.

\subsection{Insert Operation}

The \texttt{Insert} function gas benchmarks were obtained by inserting 100, 1000, 5000, and 10000 random elements into the trees.

The CMT results are presented in Table~\ref{tab:insert-keccak256} for \texttt{Keccak256} and Table~\ref{tab:insert-poseidon} for \texttt{Poseidon}. The SMT results are presented in Table~\ref{tab:smt-insert-keccak256} and Table~\ref{tab:smt-insert-poseidon}, respectively.

\begin{table}[H]
\centering
\begin{minipage}{0.45\textwidth}
\begin{tabular}{@{}cccc@{}}
\toprule
\textbf{Iterations} & \textbf{Min Gas} & \textbf{Avg Gas} & \textbf{Max Gas} \\
\midrule
100     & 97,593     & 187,682     & 301,113 \\
1,000   & 97,605     & 254,332     & 421,359 \\
5,000   & 97,605     & 286,195     & 502,851 \\
10,000  & 97,593     & 303,871     & 552,723 \\
\bottomrule
\end{tabular}
\caption{CMT gas usage with \textbf{Keccak256}}
\label{tab:insert-keccak256}
\end{minipage}
\hspace{0.05\textwidth}
\begin{minipage}{0.45\textwidth}
\begin{tabular}{@{}cccc@{}}
\toprule
\textbf{Iterations} & \textbf{Min Gas} & \textbf{Avg Gas} & \textbf{Max Gas} \\
\midrule
100     & 148,017     & 599,943     & 1,246,860 \\
1,000   & 148,017     & 883,129     & 2,068,385 \\
5,000   & 148,017     & 1,090,143   & 2,924,415 \\
10,000  & 148,017     & 1,140,019   & 2,773,845 \\
\bottomrule
\end{tabular}
\caption{CMT gas usage with \textbf{Poseidon}}
\label{tab:insert-poseidon}
\end{minipage}
\end{table}

Comparing with SMT:

\begin{table}[H]
\centering
\begin{minipage}{0.45\textwidth}
\begin{tabular}{@{}cccc@{}}
\toprule
\textbf{Iterations} & \textbf{Min Gas} & \textbf{Avg Gas} & \textbf{Max Gas} \\
\midrule
100     & 102,125     & 248,063     & 667,958 \\
1,000   & 102,137     & 294,404     & 871,190 \\
5,000   & 102,125     & 325,785     & 1,306,160 \\
10,000  & 102,125     & 339,509     & 877,732 \\
\bottomrule
\end{tabular}
\caption{SMT gas usage with \textbf{Keccak256}}
\label{tab:smt-insert-keccak256}
\end{minipage}
\hspace{0.05\textwidth}
\begin{minipage}{0.45\textwidth}
\begin{tabular}{@{}cccc@{}}
\toprule
\textbf{Iterations} & \textbf{Min Gas} & \textbf{Avg Gas} & \textbf{Max Gas} \\
\midrule
100     & 136,354     & 471,704     & 903,027   \\
1,000   & 136,366     & 620,547     & 1,522,135 \\
5,000   & 136,366     & 723,077     & 2,005,911 \\
10,000  & 136,366     & 765,205     & 2,155,222 \\
\bottomrule
\end{tabular}
\caption{SMT gas usage with \textbf{Poseidon}}
\label{tab:smt-insert-poseidon}
\end{minipage}
\end{table}

\subsection{Remove Operation}

The \texttt{Remove} function gas usage was calculated by removing all inserted elements from the trees of sizes 100, 1000, 5000, and 10000 nodes.

The CMT results are presented in Table~\ref{tab:remove-keccak256} for \texttt{Keccak256} and Table~\ref{tab:remove-poseidon} for \texttt{Poseidon}. The SMT results are presented in Table~\ref{tab:smt-remove-keccak256} and Table~\ref{tab:smt-remove-poseidon}, respectively.

\begin{table}[H]
\centering
\begin{minipage}{0.45\textwidth}
\begin{tabular}{@{}cccc@{}}
\toprule
\textbf{Iterations} & \textbf{Min Gas} & \textbf{Avg Gas} & \textbf{Max Gas} \\
\midrule
100     & 41,917     & 129,109     & 259,680 \\
1,000   & 41,917     & 178,084     & 363,816 \\
5,000   & 41,917     & 226,253     & 475,639 \\
10,000  & 41,917     & 244,545     & 522,075 \\
\bottomrule
\end{tabular}
\caption{CMT gas usage with \textbf{Keccak256}}
\label{tab:remove-keccak256}
\end{minipage}
\hspace{0.05\textwidth}
\begin{minipage}{0.45\textwidth}
\begin{tabular}{@{}cccc@{}}
\toprule
\textbf{Iterations} & \textbf{Min Gas} & \textbf{Avg Gas} & \textbf{Max Gas} \\
\midrule
100     & 92,329     & 430,727     & 1,071,617 \\
1,000   & 92,329     & 757,013     & 2,208,417 \\
5,000   & 92,341     & 889,249     & 2,143,368 \\
10,000  & 92,341     & 982,465     & 2,437,679 \\
\bottomrule
\end{tabular}
\caption{CMT gas usage with \textbf{Poseidon}}
\label{tab:remove-poseidon}
\end{minipage}
\end{table}

Comparing with SMT:

\begin{table}[H]
\centering
\begin{minipage}{0.45\textwidth}
\begin{tabular}{@{}cccc@{}}
\toprule
\textbf{Iterations} & \textbf{Min Gas} & \textbf{Avg Gas} & \textbf{Max Gas} \\
\midrule
100     & 35,029     & 131,847     & 224,916 \\
1,000   & 35,039     & 184,186     & 310,235 \\
5,000   & 35,039     & 221,294     & 443,089 \\
10,000  & 35,029     & 237,191     & 374,284 \\
\bottomrule
\end{tabular}
\caption{SMT gas usage with \textbf{Keccak256}}
\label{tab:smt-remove-keccak256}
\end{minipage}
\hspace{0.05\textwidth}
\begin{minipage}{0.45\textwidth}
\begin{tabular}{@{}cccc@{}}
\toprule
\textbf{Iterations} & \textbf{Min Gas} & \textbf{Avg Gas} & \textbf{Max Gas} \\
\midrule
100     & 35,029     & 284,457     & 466,242 \\
1,000   & 35,039     & 436,451     & 693,632 \\
5,000   & 35,029     & 544,353     & 902,871 \\
10,000  & 35,029     & 590,051     & 864,656 \\
\bottomrule
\end{tabular}
\caption{SMT gas usage with \textbf{Poseidon}}
\label{tab:smt-remove-poseidon}
\end{minipage}
\end{table}

\printbibliography

\end{document}